\documentstyle[epsfig]{basi} 
\begin{document} 
\volume{29}
\pubyear{2001}
\title[Optical and Radio observations of GRB 010222 afterglow]%
      {Optical and radio observations of the bright GRB 010222 afterglow:
       evidence for rapid synchrotron cooling?} 
\date{Received 15 April 2001; accepted 30 May 2001}
\author[R. Sagar et al]%
{R. Sagar$^{1,2}$, C. S. Stalin$^1$, D. Bhattacharya$^3$, S. B. Pandey$^1$,
\newauthor
V. Mohan$^1$, A. J. Castro-Tirado$^{4,5}$, A. Pramesh Rao$^6$, 
\newauthor
S. A. Trushkin$^7$, N.A. Nizhelskij$^7$, M. Bremer$^8$ 
\newauthor
and J. M. Castro Cer\'on$^9$\\ 
$^{1}$ State Observatory, Manora Peak, Nainital -- 263 129, India\\ 
$^{2}$Indian Institute of Astrophysics, Bangalore -- 560 034, India\\ 
$^{3}$Raman Research Institute, Bangalore -- 560 080, India\\ 
$^{4}$Instituto de Astrof\'{\i}sica de Andaluc\'{\i}a, P.O. Box 03004, 
      E-18080, Granada, Spain\\
$^{5}$Laboratorio de Astrof\'{\i}sica Espacial y F\'{\i}sica Fundamental, 
     P.O. Box 50727, E-28080, Madrid, Spain \\
$^6$ National Centre for Radio Astrophysics, TIFR, Poona University Campus, 
     Post Bag No. 3,\\ 
\phantom{X}Ganeshkhind, Pune 411 007\\
$^7$ Special Astrophysical Observatory of the Russian Academy of
   Sciences, Karanchai-Cherkessia,\\ 
\phantom{X}Nizhnij Arkhyz, 357147, Russia \\ 
$^8$ Institut de RadioAstronomie Millim\'etrique, Grenoble, France \\
$^9$ Real Observatorio de la Armada, San Fernando-Naval, 11110 C\'adiz, 
 Spain}
\maketitle 
\begin{abstract} 
We report photometric observations of the optical afterglow of GRB 010222 in 
$V$, $R$ and $I$ passbands carried out between 22--27 February 2001 at Nainital.
We determine the CCD Johnson $BV$ and Cousins $RI$ photometric magnitudes for 31
stars in the field of GRB 010222 and use them to calibrate our measurements as 
well as other published $BVRI$ photometric magnitudes of the GRB 010222 
afterglow. We construct the light curve of the afterglow emission in $B, V, R$ 
and $I$ passbands, and from a broken power-law fit determine the early and late 
time power-law flux decay indices as 0.74$\pm$0.05 and 1.35$\pm$0.04 
respectively. Steepening in the flux decay seems to have started around 0.7 day 
after the burst. Negligible Galactic extinction amounting $E(B-V) =$ 0.023 mag 
is derived in the direction of the GRB. We derive the value of the spectral 
index in the X-ray to optical region to be 0.61$\pm$0.02 and 0.75$\pm$0.02 at 
$\Delta t =$ 0.35 and 9.13 day. We attempted radio observations of the afterglow
from RATAN-600 telescope during 23--26 February 2001 and from the Giant 
Metrewave Radio Telescope on 8 March 2001, yielding upper limits of $\sim 5$ 
mJy at 3.9 GHz and $\sim 1$~mJy at 610~MHz respectively. The millimeter wave  
observations obtained during 24 and 25 February 2001 and 16 March 2001 at 93 GHz
and 230 GHz also indicate no flux detection from the afterglow. The light curve 
and the spectrum indicate that the synchrotron cooling frequency lies in the 
sub-millimeter region, which also explains the observed sub-millimeter excess. 
Attributing the observed break in the light curve to the sideways expansion of 
collimated ejecta, we estimate a jet opening angle of $\sim 2.0 n^{1/8}$~deg.
\end{abstract} 

\begin{keywords} 
Photometry -- Radio observations -- GRB afterglow -- flux decay -- spectral index
\end{keywords} 
 
\section{Introduction} 
GRB~010222 was detected simultaneously by the Gamma Ray Burst Monitor (40--700 
keV) and Wide Field Camera Unit 1 (2--28 keV) instruments aboard Beppo-SAX on 
2001 February 22 at 07:23:30 UT (Piro 2001). In the Beppo-SAX error box, the 
optical transient (OT) was detected at $\alpha_{2000} = 1 4^h 52^m 
12.^s6; \delta_{2000} = +43^{\circ} 01^{\prime} 06.^{\prime\prime}4$ 
independently by Henden (2001a) and McDowell et al. (2001). Coincident with the 
OT, Gandolfi et al. (2001) detected bright afterglow in 1.6--10 keV range about 
9 hours after the burst by Beppo-SAX TOO observations. Light curve and spectral 
index in the $X-$ray region have been presented by in 't Zand et al.
(2001). Berger \& Frail (2001) detected a radio source at 22 GHz at the 
OT location with flux of 0.7$\pm$0.15 mJy on 2001 February 22.62 UT. Fich et 
al. (2001) detected 850 $\mu$m sub-millimeter flux of 4.2$\pm$1.2 mJy at the 
same location. Near-IR $J$ and $K$ band observations of the GRB afterglow have 
been reported by Di Paola et al. (2001). A spectrum of the OT taken about 5 
hours after the burst by Garnavich et al. (2001a) reveals a blue continuum with
many superposed features corresponding to metal absorption lines at $z=1.477$,
1.157 and possibly at 0.928  (Jha et al. 2001). The existence of 3 absorption 
systems has been confirmed by low resolution spectra taken with Keck-I 10-m 
telescope (Bloom et al. 2001) and 3.58-m Telescopio Nazionale Galileo (Masetti 
et al. 2001) as well as by high resolution spectra obtained on 2001 February 
23.61 UT with Keck-II 10-m telescope (Castro et al. 2001). Absorption at 
$z=1.4768\pm0.0002$ appears to be associated with the host galaxy, while 
$z=1.1561\pm0.0001$ and $0.9274\pm0.0001$ are typical for foreground systems 
at comparable redshifts (Jha et al. 2001; Masetti et al. 2001).

The Sloan Digital Sky Survey multi-colour observations taken about 5 hours
after the burst have been reported by Lee et al. (2001). The optical and 
near-IR photometric light curves along with broad-band spectral energy 
distribution have been studied by Cowsik et al. (2001), Masetti et al. (2001) 
and Stanek et al. (2001c). Henden (2001b) presents $UBVRI$ photometric 
calibration for the GRB 010222 field region. However, the value 
of $R = 17.175\pm0.015$ mag assigned to star {\bf A} by Henden (2001b) differs 
significantly from the value of $17.42\pm0.01$ mag obtained by Valentini et al. 
(2001). We therefore imaged a number of standard regions of Landolt (1992) along
with the field of GRB 010222 for providing accurate photometric magnitudes of 
the OT. A total of 31 stars in the field have been calibrated and their standard
$BVRI$ magnitudes are given here. We present the details of our optical 
observations in the next section. Our millimeter and radio measurements and 
discussion of the light curves and other results are given in the remaining sections. 
\section { Optical observations, data reduction and calibrations }  
The optical observations of the GRB 010222 afterglow were carried out during  
2001 February 22 to 27. We used a 2048 $\times$ 2048 pixel$^{2}$ CCD system  
attached at the f/13 Cassegrain focus of the 104-cm Sampurnanand telescope of  
the State Observatory, Nainital. One pixel of the CCD chip corresponds to 
0.$^{''}$38, and the  entire chip covers a field of $\sim 13^{'} \times 13^{'}$ 
on the sky. The CCD observations of the GRB 010222 field along with Landolt 
(1992) standard regions have been carried out for calibration purposes during 
photometric sky conditions. The log of CCD 
observations is given in Table 1. In addition to these observations, several  
twilight flat field and bias frames were also observed. 
 
\begin{table} 
\caption[x]{Log of CCD photometric observations of GRB 010222 and of 
Landolt (1992) PG and SA standard fields} 
\begin{center} 
\begin{tabular}{clcl} \hline  
 Date & Field  & Filter & Exposure (in seconds) \\ \hline 
22/23 Feb 2001 & PG 1047+003 &  $V$ &   60 \\
22/23 Feb 2001 & PG 1047+003 &  $R$ &   30 \\
22/23 Feb 2001 & PG 1047+003 &  $I$ &   30 \\
22/23 Feb 2001 & PG 1323-086 &  $V$ &   40 \\
22/23 Feb 2001 & PG 1323-086 &  $R$ &   20 \\
22/23 Feb 2001 & PG 1323-086 &  $I$ &   20 \\
22/23 Feb 2001 & PG 1530+057 &  $V$ &  100 \\
22/23 Feb 2001 & PG 1530+057 &  $R$ &   40 \\
22/23 Feb 2001 & PG 1530+057 &  $I$ &   40 \\
22/23 Feb 2001 & PG 1633+099 &  $V$ &   60 \\
22/23 Feb 2001 & PG 1633+099 &  $R$ &   30 \\
22/23 Feb 2001 & PG 1633+099 &  $I$ &   30 \\
22/23 Feb 2001 & GRB 010222 & $R$ & 300$\times$3, 600$\times$2 \\ 
22/23 Feb 2001 & GRB 010222 & $V$ & 600$\times$2 \\ 
22/23 Feb 2001 & GRB 010222 & $I$ & 300$\times$2, 600$\times$2 \\ 
23/24 Feb 2001 & GRB 010222 & $R$ & 300$\times$2 \\ 
26/27 Feb 2001 & GRB 010222 & $B$ & 720 \\ 
26/27 Feb 2001 & GRB 010222 & $V$ & 600 \\ 
26/27 Feb 2001 & GRB 010222 & $R$ & 600$\times$3 \\ 
26/27 Feb 2001 & GRB 010222 & $I$ & 600 \\ 
26/27 Feb 2001 & SA 104 & $B$ & 120 \\  
26/27 Feb 2001 & SA 104 & $V$ & 60 \\  
26/27 Feb 2001 & SA 104 & $R$ & 30 \\  
26/27 Feb 2001 & SA 104 & $I$ & 30 \\  
26/27 Feb 2001 & PG 1047+003 & $B$ & 120$\times$4 \\  
26/27 Feb 2001 & PG 1047+003 & $V$ & 60$\times$4 \\  
26/27 Feb 2001 & PG 1047+003 & $R$ & 30$\times$4 \\  
26/27 Feb 2001 & PG 1047+003 & $I$ & 30$\times$4 \\  \hline 
\end{tabular} 
\end{center} 
\end{table} 

The CCD frames were cleaned using standard procedures. Image processing was done
using ESO MIDAS and DAOPHOT softwares. Atmospheric extinction coefficients were 
determined from the observations of the brightest star present in the standard  
fields and these were used in further analysis. Standard magnitudes of 27 stars 
in the standard fields were taken from Landolt (1992). They cover a wide range 
in colour ($-0.3 < (V-I) < 1.7$) as well as in brightness ($12.1 < V < 16.1$).  
The transformation coefficients were determined by fitting least square linear  
regressions to the aperture instrumental magnitudes as function of 
the standard $BVRI$ photometric indices.  The 
following colour equations were obtained for the system. 
 
\noindent $\Delta b_{CCD} = \Delta B - (0.03\pm0.01) (B-V) $  \\ 
          $\Delta v_{CCD} = \Delta V - (0.04\pm0.01) (V-R) $  \\ 
          $\Delta r_{CCD} = \Delta R - (0.02\pm0.002) (V-R) $  \\ 
          $\Delta i_{CCD} = \Delta I - (0.075\pm0.002) (R-I) $  \\ 
where $\Delta b_{CCD}, \Delta v_{CCD}, \Delta r_{CCD} $ and $\Delta i_{CCD}$ 
represent the differential instrumental magnitudes. The errors in the zero 
points are $\sim 0.02$ mag in $B, V, R$ and $I$ passbands. 
 
For increasing the photometric precision of fainter stars, the data are binned 
in $2 \times 2$ pixel$^2$ and also all the CCD images of GRB 010222 field taken 
in $R$ on 23/24 and 26/27 February 2001 are co-added in the same filter. 
 From these images, profile-fitting magnitudes are determined using DAOPHOT 
software. The standard magnitudes of the stars are determined using the above 
transformations. $BVRI$ photometric magnitudes of 31 stars in the GRB 010222 
field are listed in Table 2. Fig. 1 shows the location of the GRB 010222 
afterglow and the photometered stars on a CCD image taken by us.  

\begin{figure} 
\begin{center}
\epsfig{file=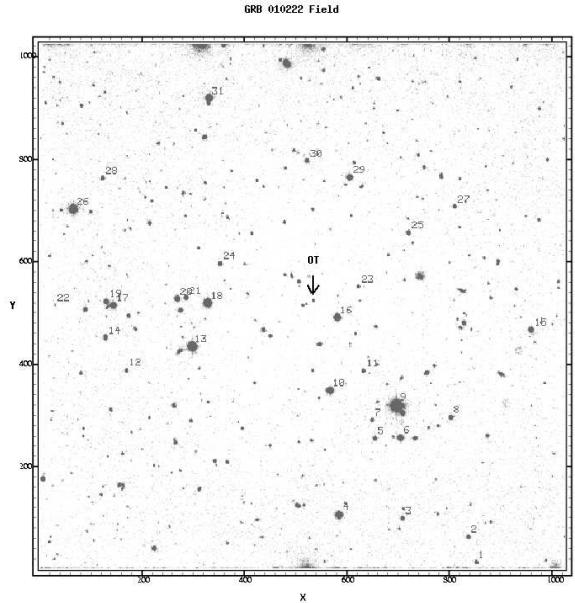,height=8cm} 
\end{center} 
\label{ident}  
\caption[x]{Finding chart for GRB 010222 field is produced from the CCD image  
taken by us on 2001 February 22.9 UT in $R$ filter with exposure  
time of 10 minutes. The optical transient (OT) and the stars with present $BVRI$
magnitudes are marked. The ($X, Y$) are the pixel coordinates and the 
corresponding sky coordinates are $\Delta\alpha$ = 767.2$\pm$0.8 + 
(0.004$\pm$0.001)$X$ - (1.038$\pm$0.001)$Y$ and $\Delta\delta$ = $-308.8\pm$0.6 
+ (0.760$\pm$0.001)$X$ + (0.001$\pm$0.001)$Y$, where $\Delta\alpha$ and 
$\Delta\delta$ are offsets in  
arcsec with respect to RA=14$^h$ 52$^m$ and Dec=43~deg respectively.}  
\end{figure} 
 
The present $BVRI$ magnitudes are compared with those determined by Henden 
(2001b) in Fig. 2. There are 26 stars common between the two data sets. They
range in brightness from $V = 14.3$ to 18.5 mag. There is no trend in $\Delta$
except a constant offset which can be understood in terms of zero-point
errors in the two photometries. We can therefore conclude that 
present and Henden (2001b) photometric calibrations are secure and calibration 
carried out by Valentini et al. (2001) may be in error.  
 
\begin{figure} 
\begin{center}
\epsfig{file=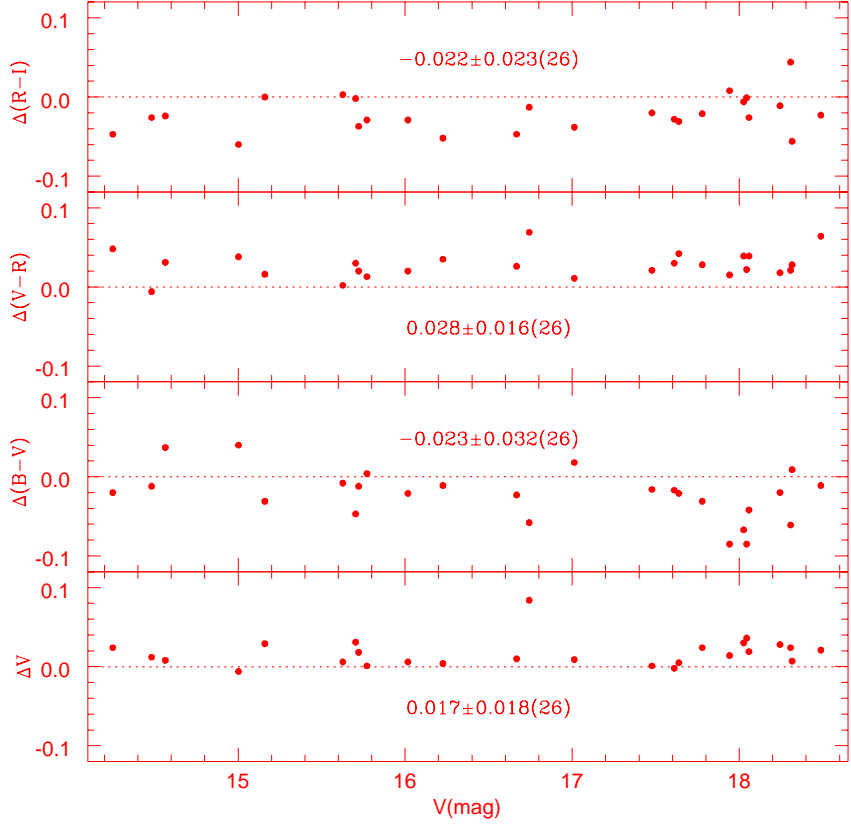,height=11cm} 
\end{center}
\label{comp}  
\caption{A comparison of our $BVRI$ photometry in the field of GRB 
010222 with data given by Henden (2001b). The differences ($\Delta$) are 
Henden minus present, plotted against the present CCD photometry. The mean
value of $\Delta$ along with number of data points (inside the bracket) are
indicated on each panel. Dotted lines represent $\Delta = 0$.}
\end{figure} 
 
The magnitude of the OT of GRB 010222 determined by us are given in Table 3.
A comparison of these with those determined independently from our CCD images
by Henden et al. (2001c) indicates agreement well within the zero-point
uncertainties of the two photometric calibrations.
The magnitudes listed in Table 2 have also been used for calibrating other 
photometric measurements of GRB 010222 afterglow published in the GCN circular 
by Holland et al. (2001), Oksanen et al. (2001), Orosz (2001), Price et al. 
(2001), Stanek \& Falco (2001), Stanek et al. (2001a, b), Valentini et al. 
(2001), Veillet (2001a, b), Watanabe et al. (2001) and  Garnavich et al. 
(2001b) by the time of paper submission. In order to avoid errors arising due 
to different photometric calibrations, we have used only those published 
$UBVRI$ photometric measurements whose magnitudes could be determined relative 
to the stars given in Table 2 or by Henden (2001b). The optical and near-IR 
magnitudes published by Cowsik et al. (2001) and Masetti et al. (2001) along 
with the Sloan Digital Sky Survey observations  presented by Lee et al. (2001) 
have also been used in our 
analysis. The photometric calibrations given by Fukugita et al. (1995) are used 
to convert the Sloan data into the Johnson $UB$ and Cousins $RI$ system. All 
these in combination with our observations provide a dense coverage of 
photometric observations of GRB 010222 afterglow in $R$ passband.
 
\begin{table}  
\caption[x]{Standard $V, (B-V), (V-R)$ and $(R-I)$ photometric magnitudes 
of the stars in the region of GRB 010222. Right Ascension ($\alpha$) 
and Declination ($\delta$) are for epoch 2000. 
Star 16 and 24 are the comparison stars A  mentioned by Henden (2001b).}
\begin{center} 
\begin{tabular}{|c|c|c|c|c|c|c|} \hline
Star&$\alpha_{2000}$&$\delta_{2000}$&$V$&$(V-R)$&$(R-I)$&$(B-V)$ \\ \hline
 1 & 14$^h$ 52$^m$ 48.$^s$00 & 43$^{\circ}$ 05$^{'}$ 07.$^{''}$2 & 18.36 & 0.82  &   0.78  &   
1.34   \\
  2  & 14 52 44.60 & 43 04 55.6 & 18.18 & 0.40 & 0.42  &   0.75   \\
  3  & 14 52 42.04 & 43 03 17.9   & 17.85 & 0.34  &   0.42   &   0.64    \\
  4  & 14 52 41.52 & 43 01 43.8   & 15.16    &   0.39  &   0.44 &   0.72   \\
  5  & 14 52 31.23 & 43 02 37.4   & 17.64    &   0.31  &   0.41 &   0.57   \\
  6  & 14 52 31.13 & 43 03 15.0   & 15.63    &   0.35  &   0.38  &   0.60   \\
  7  & 14 52 28.69 & 43 02 33.2   & 18.03    &   0.50  &   0.50  &   0.96   \\
  8  & 14 52 28.39 & 43 04 29.9   & 17.01    &   0.24  &   0.31  &   0.38   \\
  9  & 14 52 26.77 & 43 03 08.8   & 12.55    &     &     &   0.67   \\
 10  & 14 52 24.71 & 43 01 30.7   & 15.70    &   0.78 &   0.74  &   1.37   \\
 11  & 14 52 22.08 & 43 02 20.4   & 18.31    &   0.90  &  1.01 &   1.48   \\
 12  & 14 52 21.87 & 42 56 29.2   & 18.06    &   0.41  &   0.44  &   0.78   \\
 13  & 14 52 18.67 & 42 58 06.9   & 13.93    &   0.43  &   0.40  &   0.75   \\
 14  & 14 52 17.31 & 42 55 58.1   & 17.94   &    0.84  &   0.87  &   1.45   \\
 15  & 14 52 16.64 & 43 06 29.1   & 16.23    &   0.33  &   0.41  &   0.57   \\
 16  & 14 52 14.83 & 43 01 41.5   & 14.56   &   -0.15  &  -0.12  &  -0.22   \\
 17  & 14 52 13.09 & 42 56 09.8   & 15.72   &    0.48  &   0.49  &   0.93   \\
 18  & 14 52 12.81 & 42 58 29.7   & 14.48   &    0.33  &   0.35  &   0.54   \\
 19  & 14 52 12.56 & 42 55 59.3   & 16.67   &    0.40  &   0.48  &   0.75   \\
 20  & 14 52 12.22 & 42 57 44.9   & 16.02    &   0.36  &   0.39  &   0.67   \\
 21  & 14 52 12.09 & 42 57 58.0   & 17.48    &   0.58  &   0.50  &   1.05   \\
 22  & 14 52 11.69 & 42 54 57.3   & 18.60    &   0.55  &   1.97  &   1.64   \\
 23  & 14 52 10.67 & 43 02 12.7   & 18.49    &   0.54  &   0.53  &   0.99   \\
 24  & 14 52 07.54 & 42 58 48.7   & 17.61    &   0.41  &   0.44  &   0.78   \\
 25  & 14 52 03.40 & 43 03 27.1   & 17.78    &   0.84  &   0.84  &   1.44   \\
 26  & 14 52 00.09 & 42 55 10.3   & 14.25  &    0.32  &   0.39  &   0.63   \\
 27  & 14 51 56.98 & 43 04 35.9   & 18.32    &  0.24  &   0.32  &   0.38   \\
 28  & 14 51 55.95 & 42 55 54.0   & 18.24    &  0.96  &   1.12  &   1.56   \\
 29  & 14 51 55.95 & 43 02 00.2   & 15.77   &   0.34  &   0.38  &   0.61   \\
 30  & 14 51 53.71 & 43 00 56.5   & 18.04   &   0.91  &   0.96  &   1.54   \\
 31  & 14 51 45.17 & 42 58 31.7   & 15.00   &   0.31  & 0.36 & 0.50 \\ \hline
\end{tabular}
\end{center} 
\end{table} 

\begin{table}  
\caption[x]{$V, R$ and $I$ magnitudes of GRB 010222 OT along with epochs
of observations} 
\begin{center} 
\begin{tabular}{|c|c|c|} \hline
 Middle UT (day)  & filter & Magnitude$\pm$error  \\ \hline
 2001 Feb 22.911& $ R$ &  19.48$\pm$0.02 \\  
 2001 Feb 22.921& $ R$ &  19.59$\pm$0.02 \\     
 2001 Feb 22.930& $ R$ &  19.66$\pm$0.02 \\        
 2001 Feb 22.938& $ I$ &  19.07$\pm$0.03 \\        
 2001 Feb 22.949& $ I$ &  19.14$\pm$0.03 \\        
 2001 Feb 22.990& $ V$ &  20.05$\pm$0.02 \\        
 2001 Feb 22.999& $ V$ &  20.09$\pm$0.02 \\        
 2001 Feb 23.005& $ R$ &  19.80$\pm$0.02 \\        
 2001 Feb 23.010& $ R$ &  19.70$\pm$0.03 \\        
 2001 Feb 23.015& $ I$ &  19.24$\pm$0.04 \\        
 2001 Feb 23.020& $ I$ &  19.28$\pm$0.06 \\        
 2001 Feb 23.870& $ R$ &  20.77$\pm$0.35 \\     
 2001 Feb 26.911& $ R$ &  22.2~$\pm$0.10\\    \hline 
 \end{tabular}
\end{center} 
\end{table} 

\section{Optical photometric light curves} 
We have used the published data in combination with our measurements 
to study the optical flux decay of GRB 010222 afterglow in $BVRI$ photometric
passbands. Fig. 3 shows a plot of the photometric measurements as a function of 
time. The $X-$axis is log ($t-t_0$) where $t$ is the time of observation and 
$t_0$ (= 2001 February 22.308 UT) is the time of GRB trigger. All times are 
measured in unit of day.  
 
\begin{figure} 
\begin{center}
\epsfig{file=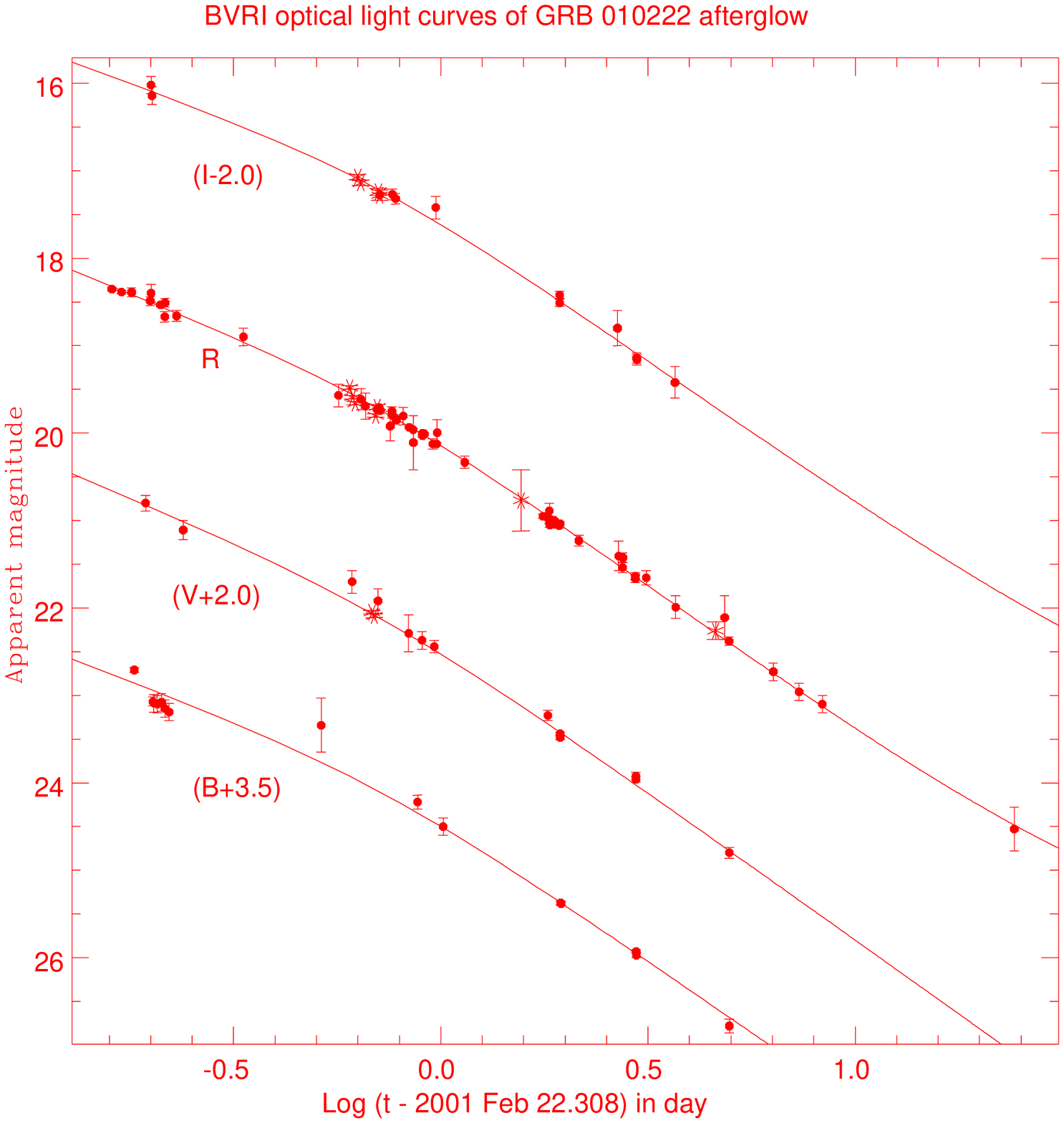,height=9.6cm} 
\end{center}
\label{light}  
\caption[x]{Light curve of GRB 010222 afterglow in optical $B, V, R$ and $I$ 
photometric passbands. Present measurements have been indicated as asterisk. 
Suitable offsets have been applied to avoid overlapping in data points of 
different passbands. Flux decay can not be fitted by a single power-law. Solid
lines represent the least square non-linear fit for jet models (see text).
In all cases, the value of sharpness is taken as 4.}  
\end{figure} 
 
The emission from GRB 010222 OT is fading in all $B, V, R$ and $I$ passbands.
As noted by Masetti et al. (2001) and Holland et al. (2001), the optical light 
curves of GRB 010222 (Fig. 3) can not be fitted by a single power-law of type 
$F(t) \propto (t-t_0)^{-\alpha}$, where $F(t)$ is the flux of the afterglow at 
time $t$ and $\alpha$ is the flux decay constant. Overall the OT flux decay 
seems to be described by a broken power-law as expected in GRB afterglows having
jet-like relativistic ejecta (Sari et al. 1999; Rhoads 1999). There is no 
signature of flattening due to contamination of background galaxy in any light 
curve though a number of galaxies are visible within 10$^{''}$ of the OT 
(cf. Garnavich et al. 2001b). As the galaxies are located more than 
4$^{''}$ away from the GRB 010222 afterglow, it appears that they have not 
contaminated the photometric measurements of the OT appreciably. However,
possibility of some light contamination can not be ruled out. We have therefore 
empirically fitted the following type of broken power-law in these light curves 
(see Sagar et al. 2000b and references therein for details) 

$F(t) = F_0[\frac{2}{{(t/t_b)^{\alpha_1 s}+(t/t_b)^{\alpha_2 s}}}]^{1/s} + F_g$,

\noindent 
where $F_g$ is the constant flux from the underlying host galaxy or other 
non-varying contaminants, $\alpha_1$ and $\alpha_2$ are asymptotic power-law 
slopes at early and late times with $\alpha_1 < \alpha_2$.  The parameter 
$s (> 0)$ controls the sharpness of the break, a larger $s$ implying a sharper 
break. $F_0$ is the flux of afterglow at the cross-over time $t_b$. In 
collimated fireball models, an achromatic break in the light curve of the 
afterglow is expected when the jet makes the 
transition to sideways expansion after the bulk Lorentz factor drops below the 
inverse of the initial opening angle of the beam (cf. Rhoads 1999). 
 
We use the dense temporal observations in $R$ to determine the parameters of 
the jet model using the above function. 
In order to avoid a fairly wide range of model parameters for a comparable 
$\chi^2$ due to degeneracy between $t_b, m_b$ and $m_g$ (magnitudes 
corresponding to $F_0$ and $F_g$ respectively), $\alpha_1$, $\alpha_2$ and $s$, 
we have used fixed values of $s$ in our analyses and find that the minimum 
value of $\chi^2$ is achieved around $s = 4$. This indicates that the observed 
break in the light curve is sharp, unlike the smooth break observed in the 
optical light curve of GRB 990510 (cf. Stanek et al. 1999; Harrison et al. 1999)
but similar to the sharp break observed in the optical light curves of GRB 
000301C (cf. Berger et al. 2000, Sagar et al. 2000b) and GRB 000926 (cf.
Harrison et al. 2001, Sagar et al. 2001). The least square fit values of the 
parameters $t_b$, $m_b$, $\alpha_1$, and $\alpha_2$ are 0.71$\pm$0.07~day, 
19.73$\pm$0.12 mag, 0.76$\pm$0.03 and 1.37$\pm$0.02 respectively in $R$ band, 
with a corresponding $\chi^2$ of 1.87 per degree of freedom ($DOF$). The
relatively large value of $\chi^2$ possibly results from slight underestimate
of systematic errors between different data sources used in this paper. The 
fit yields the constant contaminating flux to be $m_g$=26$\pm$1 mag in $R$, 
indicating that there is little contribution from the host or nearby galaxies 
(Garnavich et al. 2001b) to the photometric measurements of the GRB 010222 
afterglow till at least a week after the burst. We therefore ignore the constant
flux term in fitting the $B, V$ and $I$ light curves. The available data in 
these passbands are sparse and also insufficient for determining $t_b$ 
independently. Following Masetti et al. (2001) and Stanek et al. (2001c), we 
then assume that the break occurs at the same time in all the bands, and use the
data to determine the decay slopes. A fit with fixed $s=4$ and $t_b=0.71$~day to
the $B$ light curve yields $\alpha_1$=0.72$\pm$0.03, $\alpha_2$=1.31$\pm$0.03 
with a $\chi^2/DOF$ of 4.0; to the $V$ light curve yields $\alpha_1$=0.79$\pm$
0.05, $\alpha_2$=1.35$\pm$0.01 with a $\chi^2/DOF$ of 2.0 while such a fit in 
the $I$ light curve yields $\alpha_1=0.69\pm0.06$, $\alpha_2 = 1.33\pm0.02$ with
a $\chi^2/DOF$ of 0.9. The best fit light curves obtained in this way are shown 
in Fig. 3 for all passbands. It can be seen that our own observations follow 
the fitted curves very well. Present observations taken on 22/23 February 2001 
are around $t_b$ while those taken on other epochs fill gaps in the $R$ light
curve. They are thus valuable for the study of flux decay of the OT.

In the light of above, we conclude that the values of $t_b, \alpha_1$ 
and $\alpha_2$ derived from the $BVRI$ light curves are 0.7$\pm$0.07 day, 
0.74$\pm$0.05 and 1.35$\pm$0.04 respectively. The late time flux decay constant
is in excellent agreement with the corresponding value of 1.33$\pm$0.04 at
$X-$ray wavelength (in 't Zand 2001). They also agree very well with
the values of $t_b$ = 0.72$\pm$0.1 day, $\alpha_1 = 0.80\pm0.05$ and 
$\alpha_2 = 1.30\pm0.05$ given by Stanek et al. (2001c). However, Cowsik et al. 
(2001) and Masetti et al. (2001) have derived some what earlier time for break 
with $t_b$ = 0.43$\pm$0.1 day and 0.48$\pm$0.02 day respectively. The values
of $\alpha_1$ derived by them are 0.6$\pm$0.01 while those of $\alpha_2$ 
are 1.24$\pm$0.01 and 1.31$\pm$0.03 respectively.
\section{Radio and millimeter observations of the GRB 010222 afterglow } 
We attempted radio and millimeter observations of the GRB~010222 afterglow 
using the RATAN-600 telescope during 23--26 February 2001; the Giant 
Metrewave Radio Telescope (GMRT) on 8 March 2001 and with the millimeter Plateau
de Bure Interferometer in a compact five antenna configuration during February 
24.94 - 25.11 UT and March 16.14 - 16.40 UT.

At RATAN-600 telescope, observations were carried out at 0.96, 2.3, 3.9, 7.7,
11.2 and 21.7 GHz using the northern sector antenna with an effective 
collecting area $\sim 1000$~m$^2$.  Drift scans were obtained at the upper 
culmination of the source, and flux density calibration was performed using 
observations of 3C 286 and NGC 7027. The best sensitivity was at 3.9 GHz, with 
a noise level of 5 mJy for a single scan.  Three scans were averaged and no 
detectable signal was found at the GRB location, giving a 3-$\sigma$ upper 
limit of $\sim 10$~mJy at 3.9 GHz. 

At GMRT (see Swarup et al. 1991 for a description of the telescope) a 
$\sim 40^{\prime}\times 40^{\prime}$ field centered on the OT location was 
imaged at 617~MHz by an 8-hour synthesis on 8 March 2001.  The resolution was 
about 10$^{\prime\prime}$ and the map sensitivity was about 0.25 mJy
(the image can be viewed at 
\verb+http://www.ncra.tifr.res.in/~pramesh/images/GRB01222_8mar2001_610MHz.gif+).
No compact emission was seen at the location of the OT, and
we can place a firm upper limit of 1 mJy on the afterglow flux on this day.

At the millimeter Interferometer, weather conditions were good for observations 
at 3mm but marginal for 1mm.  The flux calibration is relative to source CRL 
618 (1.55 Jy at 3mm and 2 Jy at 1mm) and is accurate to about 10 \%. The source 
was not detected in any of the two observations within the primary beam. UV fits
on the phase center give the following upper limits (fit errors are one sigma) 
after the correction for atmospheric decorrelation.
\begin{description}
\item [(i)]During 24 February 2001 UT 22:34 to 25 February 2001 UT 2:38 the 
flux values are $-0.49\pm0.32$ mJy/beam at 93.109 GHz with a synthesized beam
of size $9.3^{''} \times 5.2^{''}$ at position angle of $-57$ degrees while the 
flux values are $-0.09\pm1.6$ mJy/beam at 232.032 GHz with beam size of 
$3.6^{''} \times 2.3^{''}$ at position angle of $-46$ degrees.
\item [(ii)] During UT 3:18 to 9:43 on 16 March 2001, the fluxes are 
$-0.42\pm0.23$ mJy/beam at 93.109 GHz and $-4.57\pm1.5$ mJy/beam at 227.239 GHz 
for the beam sizes of $9.5^{''} \times 4.9^{''}$ at position angle of 56 degrees
and $3.2^{''} \times 2.0^{''}$ at position angle of 57 degrees.
\end{description}

\section{Spectral index of the GRB 010222 afterglow} 
We have constructed the GRB 010222 afterglow spectrum at three epochs: $\Delta 
t$ = 0.35, 0.77 and 9.13 day corresponding to before, around and after $t_b$ 
respectively. The epochs were selected for the long wavelength coverage possible
at the time of $X-$ray and near-IR observations. $UJK$ data borrowed from 
Masetti et al. (2001) and $BVRI$ data along with decay slopes presented in Fig. 
3 were used to derive the fluxes at the corresponding wavelengths for the epochs
under consideration. The fluxes at 22, 220 and 350 GHz are taken from Berger \& 
Frail (2001), Kulkarni et al. (2001) and Fich et al. (2001) respectively. We 
used the reddening map provided by Schlegel et al. (1998) for estimating 
Galactic interstellar extinction towards the burst and found a small value of 
$E(B-V) = 0.023$ mag. We used the standard Galactic extinction reddening
curve given by Mathis (1990) in converting apparent magnitudes into fluxes and 
used the effective wavelengths and normalisations by Fukugita et al. (1995) for 
$UBVR$ and $I$ and by Bessell \& Brett (1988) for $J$ and $K$. The fluxes thus 
derived are accurate to $\sim$ 15\% in optical and $\sim$ 25\% in $JK$. Fig. 4 
shows the spectrum of GRB 010222 afterglow from $X-$ray to radio region. It is 
observed that at an epoch as the frequency decreases the flux increases from 
$X-$ray to radio wavelengths. For a chosen frequency interval, we describe the 
spectrum by a single power law: $F_{\nu}\propto\nu^{-\beta}$, where $F_{\nu}$ is
the flux at frequency $\nu$ and $\beta$ is the spectral index. 

\begin{figure} 
\begin{center}
\epsfig{file=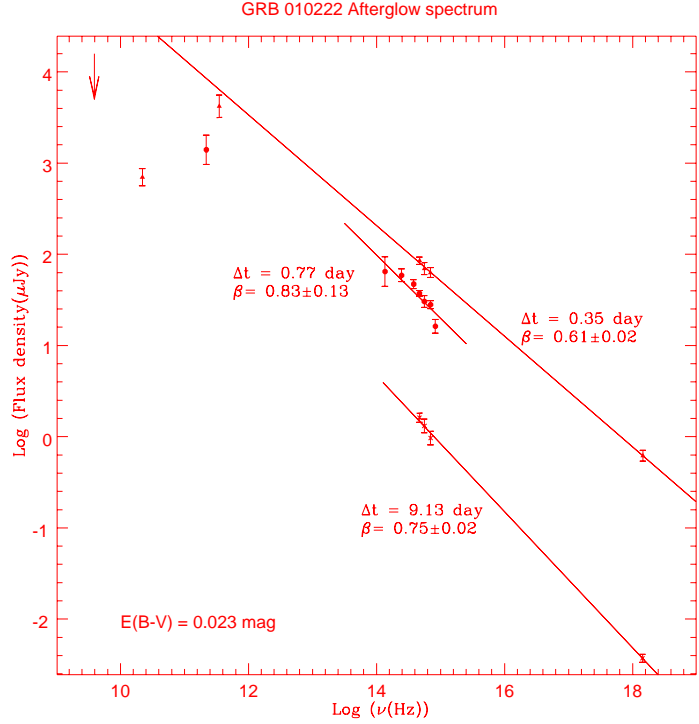,height=9.6cm} 
\end{center}
\label{spec}  
\caption[x]{The spectral flux distribution of the GRB 010222 afterglow at 
$\sim$0.35, 0.77 and $\sim$9.13 day after the burst. Arrow denotes the upper 
limit
derived by us at 3.9 GHz on 2001 February 23. The least square linear relations 
derived using fluxes at $UBVRI$ optical and $JK$ near-IR for $\Delta t = 0.77$ 
day while the fluxes at $X-$ray and optical $BVR$ for $\Delta t = 9.13$ and 0.35
day are shown by solid lines. Crosses, filled circles and triangles denote 
points corresponding to $\Delta t =$ 9.13, 0.77 and 0.35 day respectively.}  
\end{figure} 
 
The spectra at $\Delta t = 9.13$ and 0.35 day correspond to epochs of $X-$ray 
flux measurement by Harrison et al. (2001) and in 't Zand et al. (2001) 
respectively. The corresponding values of $\beta$ derived from observed flux 
values in $X-$ray and optical regions are $0.75\pm0.02$ and $0.61\pm0.02$ 
respectively. Fig. 4 indicates that these values fit the data very well. We 
therefore conclude that the values of $\beta$ have not varied significantly 
during the two epochs under discussion. When we extend the line 
corresponding to $\Delta t =$ 0.35 day to radio region, then the flux at
350 GHz fall close to this line while fluxes at $\nu <$ 220 GHz are well below 
the line indicating that the peak frequency may lie in the millimeter region of 
the spectrum. This peak frequency is thus similar to that of GRB 970508 (cf. 
Galama et al. 1998), GRB 000301C (cf. Sagar et al. 2000b) and GRB 000926 (cf. 
Sagar et al. 2001) but different from that of GRB 971214 for which the peak is 
in optical/near-IR waveband (Ramaprakash  et al. 1998). 
 
In the optical to near-IR region at $\Delta t$ = 0.77 day, the value of $\beta$ 
is $0.83\pm$0.13 which agrees within error with the values of 0.89$\pm$0.03 and 
0.90$\pm$0.03 derived about 5 hours after the burst by Jha et al. (2001) and 
Lee et al. (2001) in the optical region. All these in agreement with Masetti 
et al. (2001), perhaps, indicate no appreciable change in the optical to 
near-IR spectral slope of GRB 010222 afterglow till $\Delta t < $ 5 day. 
However, the values of $\beta$ at $\Delta t$ = 0.35 and 9.13 day derived from
observed fluxes in $X-$ray and optical region are somewhat smaller than those
derived from optical to near-IR region. The discrepancy is even larger with the 
value of 1.1$\pm$0.1 or larger quoted by Masetti et al. (2001). Fig. 4 shows 
that at $\Delta t$ = 0.77 day, the fluxes corresponding to $U, J$ and $K$ 
passbands deviate significantly from the linear relation. Both Masetti et al.  
(2001) and Lee et al. (2001) consider presence of intrinsic extinction due to
host galaxy as a possible reason for this discrepancy. This as well as the
relatively small wavelength coverage between optical to near-IR region makes
the determination of $\beta$ somewhat uncertain. On the other hand, $\beta$ 
derived from $X-$ray to optical fluxes is accurately determined due to long
wavelength coverage. We therefore consider only the values of $\beta$ derived 
at $\Delta t$ = 0.35 and 9.13 day in our further discussions.
 
\section{Discussions and Conclusions }  
GRB 010222 afterglow is rather unusual in exhibiting a relatively slow broken 
power-law decay of the light curve accompanied with flat spectral index. 
Typically the decay indices observed in such other well observed GRB afterglows 
are $\sim 1$ before the break and $\ge 2$ after the break (GRB 980519, Jaunsen 
et al. 2001; GRB 990123, Castro-Tirado et al. 1999, Kulkarni et al. 1999, Sagar 
et al. 1999; GRB 990510, Stanek et al. 1999; GRB 990705, Masetti et al. 2000a; 
GRB 991208, Castro-Tirado et al. 2001, Sagar et al. 2000a; GRB 991216, Halpern 
et al. 2000, Sagar et al. 2000a; GRB 000301C, Masetti et al. 2000b, Sagar 
et al. 2000b; GRB 000926, Sagar et al. 2001; and see also Sagar 2001).
In the case of this GRB, however, the decay index is $\sim 0.7$ before and
$\sim 1.4$ after the break. This unusual behaviour has prompted Masetti
et al. (2001) to conjecture that the break observed in GRB 010222 afterglow is
caused not by traditional sideways expansion but by the transition of the
expansion from relativistic to non-relativistic regime.

Our determination that the $X-$ray to optical spectral index of the afterglow 
has a value $\beta \le 0.75$, significantly smaller than that quoted by Masetti
et al. (2001), however, affords a different explanation of the behaviour of the 
OT. If we attribute the break in the light curve to the traditional sideways 
expansion of the jet, then the decay index after the break is expected to be 
$p$, which is the index of the power-law energy distribution of the electrons 
(Rhoads 1999). We notice then that $\beta \sim p/2$, a relation that obtains at 
frequencies above the synchrotron cooling frequency $\nu_{c}$. This would 
suggest that $\nu_{c}$ is below the optical band but above $\nu_m$ after the 
break. We then find that the same spectral regime describes the light curve well
even before the break. The observed decay index of $\alpha_1 = \sim 0.7$ fits 
nicely with the expected value of $(3p-2)/4$ (Sari et al. 1998; Rhoads 1999).

The behaviour of the light curve is therefore well modelled by having the
synchrotron cooling frequency below the optical range but above $\nu_m$ for 
nearly the entire range of observations, and attributing the break at $\sim 0.7$
days to the sideways expansion of the jet. Such a low value of cooling frequency
so early in the evolution (at $t < 0.5$~day) is certainly unusual, and
indicates the presence of a very strong postshock magnetic field.

In the above discussion we have tacitly assumed that the expressions for
light curve evolution derived in the literature (cf. Sari et al. 1998; Rhoads 
1999) apply to the case of GRB 010222 afterglow without modification. However, 
these expressions are derived assuming $p>2$, while in our interpretation of 
this afterglow $p \sim 1.2-1.5$, well below $2$. For such a flat spectral 
distribution, the upper cutoff $\gamma_{\rm u}$ of the electron energy 
distribution becomes important in the estimate of the total energy content, and 
as described in the accompanying paper by Bhattacharya (2001), a variety of 
decay laws can be expected depending on the behaviour of $\gamma_{\rm u}$ with 
time. Nevertheless, the decay laws corresponding to $p>2$ would apply also in 
this case if $\gamma_{\rm u}\propto\gamma$, the bulk Lorenz factor of the shock
(Bhattacharya 2001). Judging by the behaviour of the GRB 010222 afterglow,
this proportionality appears to have been closely followed in the present case.

An interesting consequence of $p$ being less than $2$ is that just below 
the synchrotron `cooling energy' $\gamma_{c}$ the cooling electrons pile
up, producing a `bump' in the energy distribution (cf. Pacholczyk 1970,
chapter 6), and a consequent `excess emission' around $\nu_{c}$.  Such an 
effect has indeed been noticed in the GRB 010222 afterglow at the 
sub-millimeter wavelengths (Kulkarni et al. 2001; Ivison et al. 2001; see also 
Fig.~4). The values of 350 GHz flux at $\Delta t$ = 0.23, 1.15 and 2.37 day
are 4.2$\pm$1.2, 3.6$\pm$0.9 and 4.2$\pm$1.3 mJy respectively while it becomes
undetectable at the level of 0.7$\pm$1.1 mJy on $\Delta t$ = 7.4 and 8.4 day. In
contrast to the bright sub-millimeter emission during $\Delta t <$ 3 day, the 
GRB 010222 afterglow emission is weak or undetectable at millimeter wavelengths 
(Bremer et al. 2001; Kulkarni et al. 2001; section 4). The location of $\nu_{c}$
in the sub-millimeter region would naturally explain these observations.

The break time $t_b$ can be used to determine the jet opening angle of the 
afterglow. For this, by combining the observed fluence of $(1.2\pm0.03) \times 
10^{-4}$ ergs/cm$^2$ above 2 keV (in 't Zand 2001) with the measured
redshift $z = 1.4768\pm0.0001$ (Castro et al. 2001; Jha et al. 2001), we derive 
an `isotropic equivalent' energy release $E_{\rm iso}\sim 8 \times 10^{53}$~erg
(for $H_0$ = 65 km/s/Mpc, $\Omega_0$ = 0.2 and $\Lambda_0$ = 0).
Using this value we find, from eq.~(44) of Rhoads (1999), the jet opening angle 

$\theta = 2\!\stackrel{^{\circ}}{_{\cdot}}\!0(v_{\rm l}/c)
                E_{{\rm iso},53}^{-1/8} n^{1/8}$

\noindent
where $v_{\rm l}$ is the lateral expansion speed, $E_{{\rm iso},53}$ is the
`isotropic equivalent' energy in units of $10^{53}$~erg and $n$ is the 
number density in the circumburst medium.

The peculiarity in the light curve and spectrum of GRB 010222 have been noticed
mainly due to multi-wavelength observations obtained during early times within
few hours to a day after the burst. Such multi-wavelength observations of 
recent GRB afterglows (see Sagar 2001) have thus started revealing features 
which require explanations other than generally accepted so far. We may 
therefore expect new surprises in GRB afterglows physics, once HETE and other
satellites start providing accurate positions of GRB afterglows within few 
minutes of the burst. 

\section*{Acknowledgements}
This research has made use of data  
obtained through the High Energy Astrophysics Science Archive Research Center  
Online Service, provided by the NASA/Goddard Space Flight Center. Our 
special thanks to Scott Barthelmy and Paul Butterworth for running the
GCN circular and notice services. We are grateful to Dr. L. Piro for providing
$X-$ray flux value before publication. S.A.T. and N.A.N. are thankful to
Prof. Yu.N. Parijskij for including the observations of GRB 010222 
in his program at RATAN-600 radio telescope.  The GMRT is the result
of dedicated and untiring effort of many people at the National Centre
for Radio Astrophysics, Pune and other associated institutions.

\end{document}